\documentclass[useAMS,usenatbib]{mn2e}
\usepackage{graphicx}
\newcommand{\etal}{{\it et al.} }

\title[Black hole mass, Host galaxy classification and AGN activity]{
Black hole mass, Host galaxy classification and AGN activity }

\author[B. McKernan, K.E.S. Ford \& C.S.Reynolds]{B. McKernan$^{1,2}$\thanks{E-mail:bmckernan at amnh.org (BMcK)}, K.E.S. Ford$^{1,2}$ \& C.S. Reynolds$^{3}$\\
$^{1}$Department of Science, Borough of Manhattan Community College, City University of New York, New York, NY 10007\\
$^{2}$Department of Astrophysics, American Museum of Natural History, New York, NY 10024\\
$^{3}$Department of Astronomy, University of Maryland, College Park, MD 20742\\}
\begin{document}

\date{Accepted. Received; in original form}

\pagerange{\pageref{firstpage}--\pageref{lastpage}} \pubyear{2008}

\maketitle

\label{firstpage}

\begin{abstract}
We investigate the role of host galaxy classification and black hole mass 
($M_{BH}$) in a heterogeneous sample of 276 mostly nearby ($z<0.1$) X-ray and 
IR selected AGN. Around 90$\%$ of Seyfert 1 AGN in bulge-dominated host 
galaxies (without disk 
contamination) span a very narrow range in the observed 12$\micron$ to 2-10keV 
luminosity ratio ($1<R_{IR/X}<7$). This narrow dispersion 
incorporates \emph{all} possible variations among AGN central engines, 
including accretion mechanism and efficiency, disk opening angle, orientation 
to sightline, covering fraction of absorbing material, patchiness of X-ray 
corona and measured variability. As a result, all models of X-ray and IR 
production in AGN are very strongly constrained. Among Seyfert 1 AGN, median 
X-ray and 
IR luminosities increase with black hole mass at $>99\%$ confidence. Using 
ring morphology of the 
host galaxy as a proxy for lack of tidal interaction, we find that AGN 
luminosity in host galaxies within 70Mpc is independent of host galaxy 
interaction for $\sim$ Gyrs, suggesting that the timescale of AGN activity 
due to secular evolution is much shorter than that due to tidal 
interactions. We find that 
LINER hosts have lower 12$\micron$ luminosity than the median $12\micron$ 
luminosity of normal disk- and bulge-dominated galaxies which may represent 
observational evidence for past 
epochs of feedback that supressed star formation in LINER host galaxies. We 
propose that nuclear ULXs may account for the X-ray emission from LINER 2s 
without flat-spectrum, compact radio cores. We confirmed the robustness of 
our results in X-rays by comparing them with the 14-195keV 22-month BAT 
survey of AGN, which is all-sky and unbiased by photoelectric absorption.

\end{abstract}


\section{Introduction}
\label{sec:intro}
Galactic nuclei generally considered to be 'active' have bolometric luminosities of 
$\sim 10^{-2}$ to $\sim 10^{4}$ times their host galaxy luminosity and are 
powered by accretion onto a supermassive black hole. The 
fundamental parameters that determine accretion onto the central black hole 
should be: (1) the black hole 
mass, (2) the amount of gas and dust in the galactic nucleus, (3) a mechanism 
to drive material in the galactic nucleus onto the black hole and (4) outflows
 as a result of accretion, which may affect (2) and (3) as feedback. While 
most galactic nuclei in the local Universe are believed to host very large 
mass black holes, only $\sim 0.01-1\%$ of these nuclei are highly luminous 
\citep[e.g.][]{b49,b13}. By contrast, low luminosity nuclei are far more 
common, accounting for $\sim 1/3$ 
of all galactic nuclei \citep{b95}. Therefore some combination of the 
parameters 
determining accretion conspires to keep very luminous galactic nuclei 
relatively rare, while also providing a very wide range of observed activity 
at lower luminosities. 

Black hole mass is the simplest parameter involved in accretion. The greatest 
activity (or highest bolometric luminosity) in a galactic nucleus occurs when 
a black hole accretes mass quickly. Black hole mass correlates
 with several properties of the central regions of 
the host galaxy, including bulge stellar velocity dispersion ($\sigma_{\ast}$),
 bulge mass and bulge 
luminosity (see e.g. \citet{b39} and references therein). This implies that 
the largest mass black holes live in galaxies with large bulges. If most
galactic bulges grew early in the Universe \citep[e.g.][]{b41}, then black 
holes grew fastest at that time and so activity in 
galactic nuclei may have been greatest at high redshift. However, periods of 
intense activity in a galactic nucleus could also arise at lower redshift, due
 to mergers between gas-rich galaxies \citep[e.g.][]{b43} or when a large 
reservoir of cold gas builds up in the galactic disk \citep{b24,b40}.

The raw material for accretion onto the black hole is gas and dust in the 
galactic nucleus. How much gas and dust there is in the galactic nucleus 
depends on the local rate of formation of massive stars \citep{b25} and on
 mechanisms driving gas and dust into the nucleus from elsewhere in the 
galaxy. Mechanisms 
driving material into the galactic nucleus from 
the outside could be internal or external to the galaxy. Internal mechanisms 
involve bars \citep[e.g.][]{b33} or more generally, internal (secular) disk 
driven evolution \citep{b24}; external mechanisms involve tidal disruptions or
 mergers \citep[e.g.][]{b41} or nuclear bombardment \citep{b60}. Not all of 
the material in the galactic nucleus 
needs to come from the rest of the galaxy. Material that gains angular 
momentum close to the accreting supermassive black hole can flow outwards 
to be recycled in the surrounding galactic nucleus 
\citep[e.g.][]{b27,b1,b44,b8}. Indeed if the feedback from the 
accreting black hole is powerful enough it could disrupt star formation in 
the galactic nucleus and beyond \citep{b1}. Outflows from the accreting 
black hole into the
 surrounding galactic nucleus could end up terminating inflows onto the 
black-hole itself, which leads to a picture of galactic nucleus activity
 as self-regulating \citep[e.g.][]{b38}. 

Isolating the fundamental parameters that determine the mass accretion rate 
onto supermassive black holes is a major observational problem. Many active 
galactic nuclei 
(AGN) are shrouded by obscuring material \citep{b2}, although the obscuration 
can be complicated \citep[e.g.][]{b4,b3}. AGN are 
mostly distant enough that broad-band observations include host galaxy 
luminosity contributions (e.g. from hot diffuse gas, X-ray binaries and 
ultra-luminous X-ray sources in the X-ray band alone). One approach to 
solving this difficult observational problem is 
to compare broadband luminosities in AGN with fundamental accretion 
parameters, such as black hole mass, or simple observables such as host galaxy
 classification, which may be related to fundamental accretion parameters. 

In this work, we investigate the connection between black hole mass, host
 galaxy classification and the observed IR and X-ray luminosities of a 
heterogeneous sample of 276 AGN (mostly from \citep{b99}). In 
section~\ref{sec:sample} we investigate the connection between AGN 
luminosity, black hole mass and host galaxy classification. In 
section~\ref{sec:hockey} we discuss the AGN luminosity distribution and in 
section~\ref{sec:group2} we discuss the highly heterogeneous, non-Seyfert 1 
AGN in our sample. In 
section~\ref{sec:rings}, we discuss the importance of 
ringed morphology in host galaxies and the implications for AGN activity. We 
discuss 
issues of bias and completeness in our sample as well as the reliability of 
our conclusions in section~\ref{sec:bias} and in section~\ref{sec:conclusions}
 we summarize our conclusions.

\section{A heterogeneous sample of AGN}
\label{sec:sample}
In \citet{b99}, for the first time we compared 
observed 2-10keV X-ray and $12-100\micron$ infra-red luminosities in a large, 
heterogeneous sample of 245 AGN in the literature. Heterogeneous surveys such 
as this one include multiple biases and for a discussion of these, see 
\citet{b99} and Section~\S\ref{sec:bias} below. In some sources, the relative 
contribution of an AGN to the IR and X-ray luminosity of a LINER, starburst or 
ULIRG can be debated. Indeed multiple activity classifications are often given
 to the same source. In our sample, since we do not know the relative 
contribution of the AGN \emph{a priori}, we do not exclude these objects. 
Furthermore, we avoid model assumptions about specific galactic nuclei or AGN 
types by simply using observed luminosities (host galaxy contamination 
included). Since AGN classification is complicated, in \citet{b99} we 
simply divided our sample in two: Group 1 AGN are Seyfert 1.X AGN (X=0-9) or 
QSOs. Group 2 AGN are everything else (Seyfert 2s, LINERs, low luminosity AGN, 
Starburst/AGN, cross-classified AGN). Table~\ref{tab:sample} lists some 
of the AGN in 
our sample, with the full sample in a machine-readable table online. 
Our key finding in 
\citet{b99} was that, once we removed highly beamed sources, the 
\emph{dispersion} in the ratio of observed IR to X-ray luminosities in Group 1 
AGN is narrow. A narrow dispersion in physical properties permits the 
establishment of strong constraints on the physical properties of 
the AGN, independent of assumptions about the central engine (e.g. orientation
 to sightline, clumpy torus, nature of accretion flow, extent or patchiness of
 X-ray corona and even AGN variability). For this work, we found additional 
AGN in the literature \citep{b91,b66,b53,b56,b45,b57} to add to the sample in 
\citet{b99}. 

\begin{table*}
\begin{minipage}{120mm}
\caption{A guide to the AGN in our sample. The full sample is listed in a 
machine readable table online. Column 1 is AGN name, column 2 is the 
luminosity distance (Mpc) listed in NED. Columns 3 and 4 are mean observed 
2-10keV X-ray 
flux ($\times 10^{-11}$ ergs $\rm{cm}^{-2} \rm{s}^{-1}$) and 12$\micron$ flux 
(Jy) respectively. Column 5 is host galaxy classification 
(from NED notes and \citep{b71}) and Column 6 is activity classification 
according to NED and \citep{b54} with (in brackets) Group 1 or 2 
identification. Activity classification is denoted according to 
\citep{b54} except: we write Sy3 as L(LINERS) with S3b,S3h denoted by L1 
and L2 respectively and S1n are written as NLSy1 (narrow-line Sy 1). Other 
classifications include SB=starburst,U=ULIRG. Column 7 is 
central black hole mass estimate 
and Column 8 is references for data. References: N=NED, 1= \citet{b91}, 
2=\citet{b56}, 3=\citet{b28}, 4=\citet{b30}, 5=\citet{b45}, 6=\citet{b53}, 
7=\citet{b55}, 8=\citet{b59}, 9=\citep{b66}, 10=\citep{b67},11=\citep{b68},
12=\citep{b69}, 13=\citep{b72}. 
\label{tab:sample}}
\begin{tabular}{@{}lrrrrrrr@{}}
\hline
Name & $L_{d}$ & $\overline{F}_{x}$ & $\overline{F}_{ir}$ & Host & AGN(Group) & log($M_{BH}$) & Ref.\\
    & &  &  & & & &\\
\hline
PG 1404+226 & 437 & 0.01  & 0.03& -           &NLSy1 (1)  & 6.70 &N, \\ 
RE 1034+396 & 184 & 0.10   & 0.13 & SBa;compact & NLSy1(1)  & 6.81 &N, \\
Ton S180    & 263 & 0.46  & 0.12 & SABa        & Sy1.2(1)  & 6.85 &N,1 \\
Akn 120     & 138 & 3.30   & 0.30 & SAb pec     & Sy 1(1)   &  -    &N, \\
Mkn 1040    & 66.1& 2.00  & 0.61 & SAbc int    & Sy 1.0(1) & 6.36 &N,2\\
\hline
\end{tabular}
\end{minipage}
\end{table*}

\begin{figure}
\includegraphics[height=3.35in,width=3.35in,angle=-90]{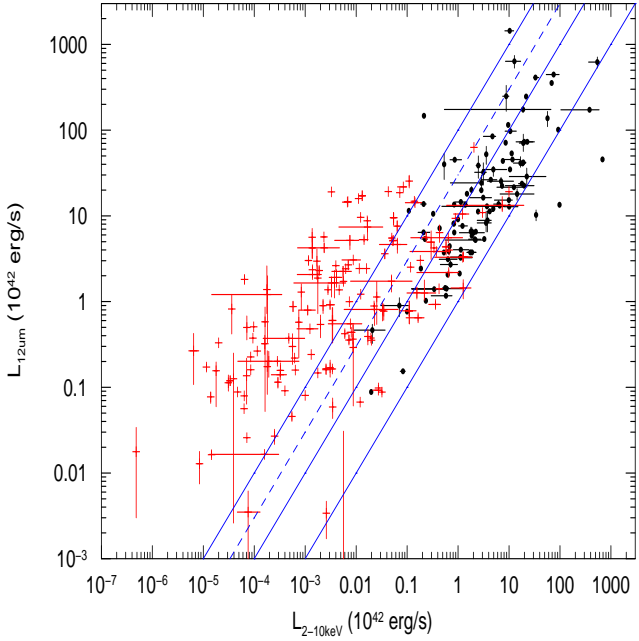}
\caption{Mean observed 12$\micron$ IR luminosity 
versus mean observed 2-10keV luminosity for the AGN in our sample (listed in 
Table~\ref{tab:sample}), excluding 
sources classified as highly beamed. Group 1 AGN
 are denoted by filled-in circles and Group 2 AGN by crosses. Error bars 
where indicated correspond to the range of multiple luminosity measurements. 
 AGN observed once in a waveband are not given error bars in that waveband 
since measurement error is not representative of AGN 
variability. Solid lines indicate constant luminosity ratios of 1,10 and 100.
 Dashed line 
indicates constant luminosity ratio of 30. AGN whose classification is 
jet-dominated (e.g. BL Lacs) are excluded. Note: Cen A has a 
value of $R_{IR/X}=15.7$ not $0.157$ as plotted in 
\citet{b99}.
\label{fig:hockey}}
\end{figure}

In Fig.~\ref{fig:hockey} 
we plot the mean observed 12$\micron$ luminosity versus mean 
observed 2-10keV luminosity for the AGN in our sample that are not 
jet-dominated. Black circles correspond to Group 1 AGN and red crosses 
correspond to Group 2 AGN. Error bars correspond to the range of observed 
luminosities for a given AGN. These error bars (where present) also reveal 
observer bias in our sample, since these are the AGN that have been
 observed multiple times. From Fig.~\ref{fig:hockey}, the AGN distribution is 
generally 'hockey-stick' shaped. Even including variability, the Group 1 AGN 
lie in a fairly tight 
range of luminosity ratio (90$\%$ lie in the range $R_{IR/X}=[1,30]$), forming
 the (black) 'stick'. The Group 2 AGN overlap with the Group 1 AGN but then 
flare out 
to much larger values of $R_{IR/X}$, forming the (red) 'blade' of the hockey 
stick. 
In section \S~\ref{sec:hockey} below we shall discuss the significance of the 
hockey-stick shape of the AGN distribution. In this section we shall 
investigate the role played by black hole mass and host galaxy type in the 
observed IR and X-ray luminosities of the AGN in our sample. Since the Group 1 
AGN contain less variation in the type and range of activity, we shall 
investigate these objects first.

\subsection{Group 1 AGN}
\label{sec:gr1}
To start, we shall investigate the relationship between black hole mass and 
observed IR and X-ray luminosity of Group 1 AGN. Studies of stellar and gas 
kinematics in the centers of nearby galaxies 
(and indeed our own) reveal the presence of supermassive black holes 
\citep{b9}. Estimates of the central black hole mass from kinematics correlate
 with host galaxy properties, such as the velocity dispersion of stars in the 
bulge as well as bulge luminosity \citep[see e.g.][and references 
therein]{b39}. It has therefore become possible to measure (with large 
uncertainty) central black hole masses in AGN (and in normal galaxies) based on
 observed values of $\sigma_{\ast}$ \citep[e.g.][]{b91} and bulge luminosity 
\citep[e.g.][]{b56}. Black hole mass in AGN can also be estimated via 
reverberation mapping \citep[e.g.][]{b14,b58}, but see also criticism of the 
claimed accuracy \citep{b20}.

\begin{figure}
\includegraphics[height=3.35in,width=3.35in,angle=-90]{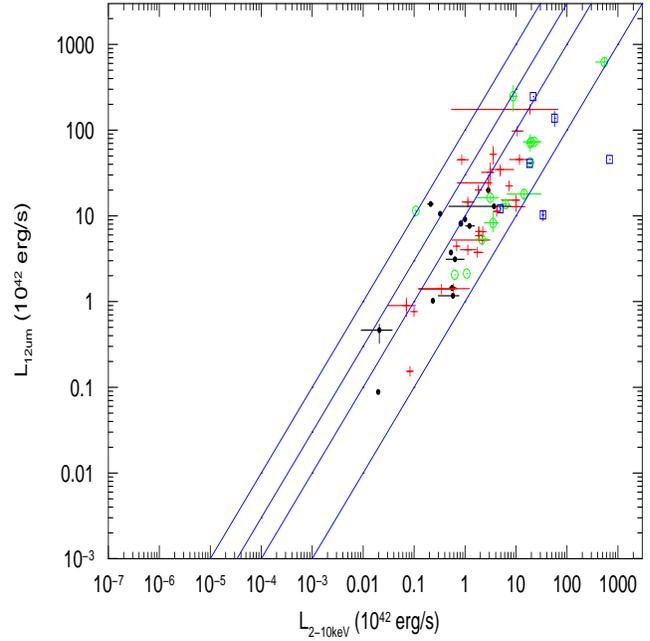}
\caption{As Fig.~\ref{fig:hockey} except for the Group 1 AGN in our sample 
with 
estimates of black hole mass in the literature. Black filled-in circles, red 
crosses, green open circles and blue open squares denote $M_{BH}<10^{7},
10^{7-8},10^{8-9},>10^{9} M_{\odot}$ respectively. Error bars are as in 
Fig.~\ref{fig:hockey}. 
\label{fig:mass1}}
\end{figure}

Of the AGN in our sample, some 154 had estimates of black hole mass in the 
literature (from references to Table~\ref{tab:sample}). Where there are 
significant differences between 
black hole masses estimated for the same AGN, we used the 
value estimated using $\sigma_{\ast}$, since there is systematic error in 
bulge luminosity estimates in spiral galaxies, due to bulge-disk decomposition 
techniques \citep[e.g.][]{b19}. So our $M_{BH}$ estimates will be 
biased in favour of those AGN in host galaxies where $\sigma_{\ast}$ can be 
accurately deduced. In Figure~\ref{fig:mass1} we plot the 
mean observed 12$\micron$ IR luminosity versus the mean observed 2-10keV X-ray
 luminosity for Group 1 AGN, coloured according to 
estimated black hole mass.  From Fig.~\ref{fig:mass1}, average X-ray and IR 
luminosities tend to increase with black hole mass, although for a given 
black hole mass, luminosities seem to span two or three orders of magnitude 
($\sim 10^{-4}-10^{-1} L_{edd}$). However there is 
considerable uncertainty in the estimated values of black hole mass in the 
literature. Different
 methods used to estimate black hole masses in the same AGN generally agree 
within a factor of few, but can differ by more than an order of magnitude 
\citep[e.g.][]{b91,b56}. Reverberation mapping \citep[e.g.][]{b58} may 
involve 
uncertainties a factor of $\sim 3$ greater than estimated by authors 
\citep{b20}. Intrinsic dispersion in the correlation with $\sigma_{\ast}$ as 
well as 
bulge-disk decomposition errors \citep{b19} are further sources of 
uncertainty in 
black hole mass estimates via the $M-\sigma_{\ast}$ relation and bulge 
luminosity respectively. 

In order to reduce the effect of relatively large uncertainties in 
measurements of black 
hole masses, we simply divided our sample of AGN into low mass ($<10^{7} 
M_{\odot}$) and high mass ($>10^{8} M_{\odot}$) populations. Estimated 
uncertainties in the measured mass of black holes generally do not exceed a 
factor of 10, so this strategy should enable us to assemble two distinct 
populations with few or no overlapping members. The T-test reveals that low and 
high mass Group 1 AGN do not have significantly different mean ratios of 
IR to X-ray luminosity ($R_{IR/X}$) at a confidence level of $\sim 99\%$. 
So Group 1 AGN appear to maintain a narrow range of $R_{IR/X}$ across 3-4 
orders of magnitude of $M_{BH}$.

\begin{figure}
\includegraphics[height=3.35in,width=3.35in,angle=-90]{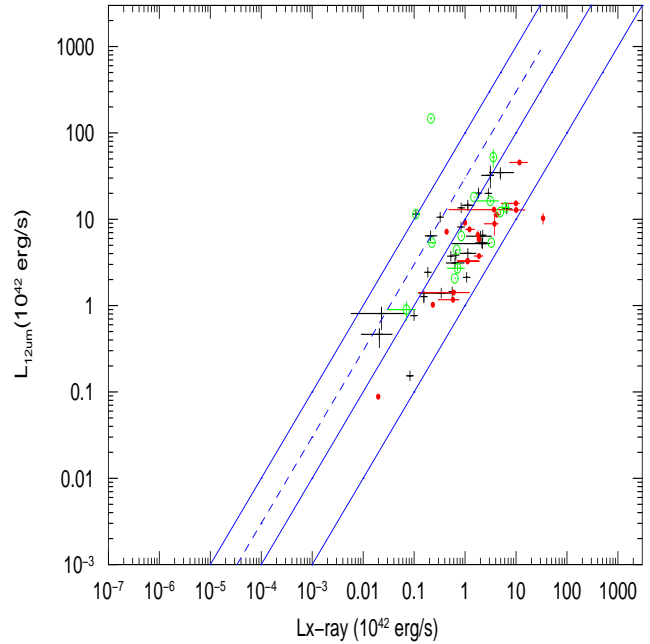}
\caption{As Fig.~\ref{fig:hockey} except for the Group 1 AGN in our sample 
with $z<0.1$ and host galaxy classifications.
 Black crosses denote AGN with disk-dominated hosts, red filled-in circles 
denote AGN with bulge-dominated hosts and green open circles denote AGN with 
disrupted hosts. Errors are as in Fig.~\ref{fig:hockey}.
\label{fig:class1}}
\end{figure}

Now we investigate the role of host galaxy classification in the 
observed IR and X-ray luminosities of Group 1 AGN. We follow other galaxy 
surveys by simply splitting AGN host galaxies into 3 
groups: elliptical and lenticular (bulge-dominated), spiral (disk-dominated) 
and disrupted (peculiar morphology due to merging or tidal interaction). We 
separate bulge-dominated from disk-dominated hosts at Hubble stage 
$T=-0.5$ (e.g. \citet{b35}), so that a galaxy classed as SA0+ ($T=-1$) 
for example, is bulge dominated ('early') but SA0/a ($T=0$) is disk-dominated 
('late'). We are therefore sensitive to classification 
error among (S0+,S0/a) galaxies. However 
galaxies in these two stages are $<10\%$ of our sample. Table~\ref{tab:host} 
lists the number of AGN in our sample 
according to NED classification of their hosts as bulge-dominated, 
disk-dominated or disrupted. 

\begin{table}
\begin{minipage}{60mm}
\caption{AGN in our sample divided according to host galaxy classification 
from NED. Bulge-dominated hosts correspond to Hubble stages $T<-0.5$ and
 disk-dominated hosts correspond to Hubble stages $T>-0.5$. Host galaxies with
 disrupted morphologies (peculiar/interacting) are 
listed separately. $^{a}$ 3 Group 1 AGN host galaxies were classified as 
'Compact'. 
\label{tab:host}}
\begin{tabular}{@{}lrrrr@{}}
\hline
                      & bulge$^{a}$ & disk & disrupted &Total\\
\hline
Group 1   & 22 &28 &16 &66\\
Group 2               & 40&72&32 &143\\
\end{tabular}
\end{minipage}
\end{table}

In Figure~\ref{fig:class1} we plot the mean observed 12$\micron$ IR 
luminosity 
versus mean observed 2-10keV X-ray luminosity for the Group 1 AGN from 
Table~\ref{tab:host}, colour coded according to host galaxy type. 
Note that there are considerably more galaxies in our sample with host galaxy 
classifications (209) than with black hole mass estimates (154). From 
Fig.~\ref{fig:class1}, those AGN in hosts that are bulge-dominated 
(red filled-in circles) tend to pile up closer to the $R_{IR/X}=1$ line than 
the disk-dominated AGN (black crosses). The 
mean (median) value of $R_{IR/X}$ for disk-dominated Group 1 AGN (black 
crosses) in Fig.~\ref{fig:class1} is a factor of 3.4(2.5) larger than for 
bulge-dominated 
Group 1 AGN (red filled-in circles) and the two populations have different 
mean ratios of $R_{IR/X}$ at a confidence level $>99\%$ from the RS-test. 

One obvious source of the greater dispersion among disk-dominated Group 1 AGN 
is emission from unresolved galactic disks. Note that host galaxy IR emission 
will be unresolved by the $\sim 30''$ beamwidth of IRAS for most AGN in our 
sample. However, ground-based narrow beam (6-9$''$) 
$10.6\micron$ measurements of 10 nearby AGN observed by IRAS, reveal that host
 galaxy disks may account for around half the IRAS flux measurement 
\citep{b12}. We found that the median IR luminosities of disk-dominated Group 
1 AGN is $\sim 10^{42}$ erg $\rm{s}^{-1}$ greater than the median IR 
luminosity of bulge-dominated Group 1 AGN. This IR excess is presumably due to
 the unresolved galactic disk (see also the discussion in \S~\ref{sec:hockey} 
below). However, we also found that the median 2-10keV X-ray luminosity in 
disk-dominated Group 1 AGN is a factor of 2 lower than in bulge-dominated 
Group 1 AGN. This median X-ray deficit ($\sim 10^{42}$ erg $\rm{s}^{-1}$) 
may be due to a greater covering fraction of dust and gas among AGN in 
disk-dominated hosts. The heating caused by the absorbed or scattered X-rays 
will emerge isotropically from the covering dust, which should in turn 
contribute to the observed IR excess.

So there are two components of the greater scatter among disk-dominated Group 
1 AGN. First is an average IR excess of $\sim 10^{42}$ erg $\rm{s}^{-1}$ 
probably due to contamination by the unresolved host galaxy. Second is an 
X-ray deficit of $\sim 10^{42}$ erg $\rm{s}^{-1}$ possibly due to absorption 
by a greater covering fraction of gas and dust, which would contribute to IR 
excess. Therefore true dispersion in $R_{IR/X}$ observed in Group 1 AGN, is 
best represented by the bulge-dominated Group 1 AGN. 
The dispersion among our sample of bulge-dominated Group 1 AGN alone is 
$R_{IR/X}=[1,16]$ but $\sim$90$\%$ (14/16) lie within the range 
$R_{IR/X}=[1,7]$, which is a remarkably narrow dispersion considering that it
 includes variation in all central engine model parameters.

\subsection{Group 2 AGN}
\label{sec:gr2}
\begin{figure}
\includegraphics[height=3.35in,width=3.35in,angle=-90]{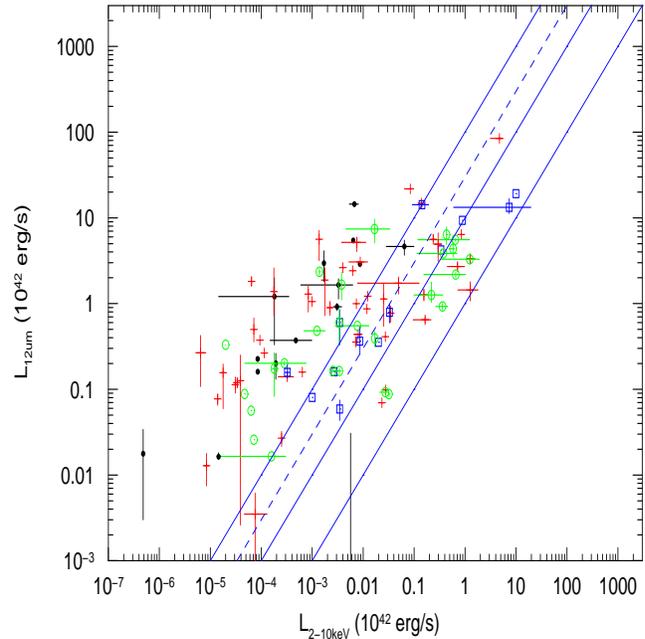}
\caption{As Fig.~\ref{fig:mass1} except for Group 2 AGN.  
\label{fig:mass2}}
\end{figure}

Here we investigate the role that black hole mass and host 
galaxy classification play in the observed X-ray and IR luminosities of Group 
2 AGN, the 'blade' of the hockey-stick distribution in Fig.~\ref{fig:hockey}. 
In Figure~\ref{fig:mass2} we plot the 
mean observed 12$\micron$ IR luminosity versus the mean observed 2-10keV X-ray
 luminosity for Group 2 AGN, coloured according to estimated black hole mass. 
Clearly Group 2 AGN behave differently to the Group 1 AGN. Most Group 2 AGN 
have low values of $L_{X}$($<10^{42}$erg $\rm{s}^{-1}$), but $L_{IR}$ can 
range from low ($< 10^{40}$erg $\rm{s}^{-1}$) to ultra-luminous 
($> 10^{43}$erg $\rm{s}^{-1}$). From Fig.~\ref{fig:mass2}, the ratio of IR to 
X-ray luminosities ($R_{IR/X}$) broadly tends to 
\emph{decrease} with black hole mass, going from $<10^{8}M_{\odot}$ 
(black filled in circles and red crosses) to $10^{8-9}M_{\odot}$ (green open 
circles) to $>10^{9}M_{\odot}$ (blue open squares). However, we saw above in 
\S\ref{sec:gr1} that 
disk-dominated hosts could contribute significantly to the observed IR 
luminosity. So host galaxy classification will be important for the 
interpretation of our results.

\begin{figure}
\includegraphics[height=3.35in,width=3.35in,angle=-90]{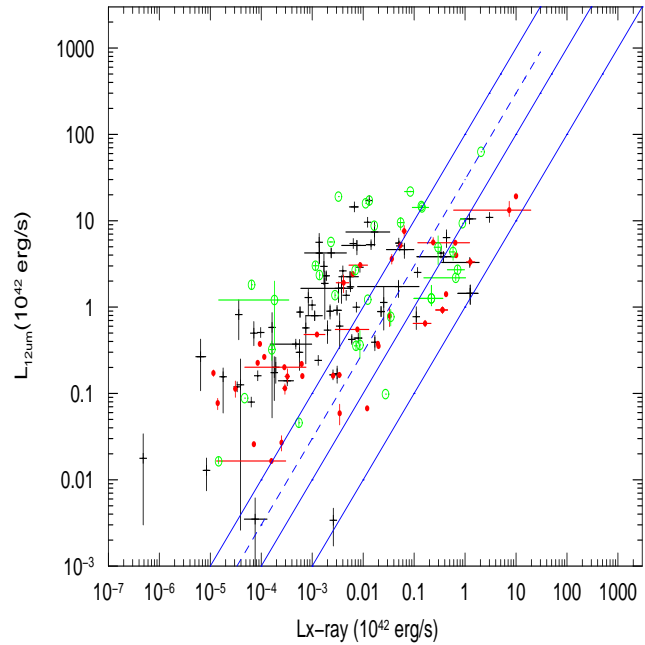}
\caption{As Fig.~\ref{fig:class1} except for Group 2 AGN.
\label{fig:class2}}
\end{figure}

In Figure~\ref{fig:class2} we plot the mean observed 12$\micron$ IR luminosity 
versus mean observed 2-10keV X-ray luminosity for the Group 2 AGN from 
Table~\ref{tab:host}, plotted according to host galaxy type. 
Those Group 2 AGN in disk-dominated hosts (black crosses) tend to have higher 
median $L_{IR}$ than Group 2 AGN in bulge-dominated hosts 
(red filled-in circles) at low $L_{x}(<10^{40} \rm{erg} \rm{s}^{-1})$. 
This is consistent with a 
larger contribution to the observed 
IR luminosity from unresolved disk-dominated hosts (as we found for 
Group 1 AGN in Fig.~\ref{fig:class1}), although this may not be the whole 
story (see \S~\ref{sec:hockey} below). The scatter in $L_{X}$ in 
Fig.~\ref{fig:class2} is comparable for both bulge- and disk-dominated Group 
2 AGN (unlike Group 1 AGN). Many of the Group 2 AGN with 
the highest $L_{IR}$ in Fig.~\ref{fig:class2} live in disrupted hosts 
(green open circles). This is probably due to an increase in star formation 
in the host galaxy rather than 
increased AGN activity. Otherwise we should observe the same trend in 
Fig.~\ref{fig:class1} among 
Group 1 AGN in disrupted hosts. AGN on the boundary between 
Group 1 and Group 2 classifications, i.e. cross-classified Group 2 AGN that 
include a Sy1.X classification appear to be dominated by the Sy1.X component 
in bulge-dominated hosts, since their median $R_{IR/X} \sim 9$ is consistent 
with that of Group 1 AGN.

\begin{figure}
\includegraphics[height=3.35in,width=3.35in,angle=-90]{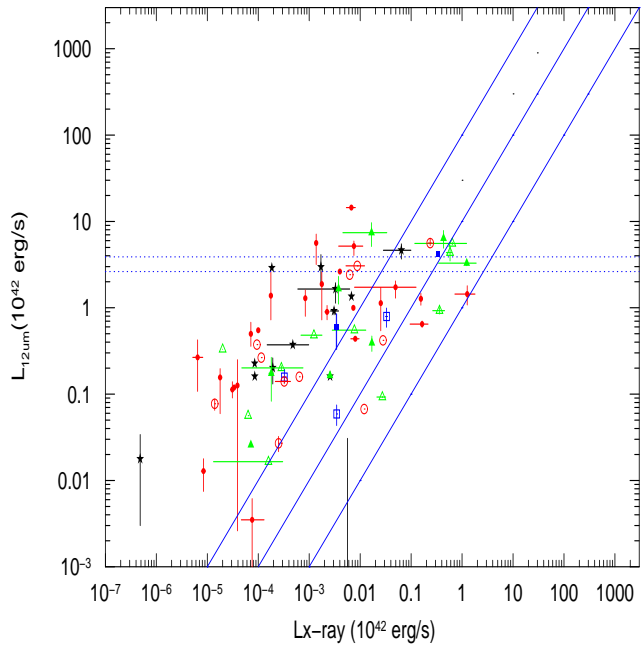}
\caption{Combined information from Fig.~\ref{fig:mass2} and 
Fig.~\ref{fig:class2} respectively. Black filled-in stars denote AGN with 
$M_{BH}<10^{7}M_{\odot}$ and disk-dominated hosts. In red are AGN with 
$M_{BH}=10^{7-8}M_{\odot}$ where filled-in and open circles denote disk and 
bulge dominated hosts respectively. In green are AGN with $M_{BH}=10^{8-9} 
M_{\odot}$ where filled-in and open triangles denote disk and bulge-dominated 
hosts respectively. In blue are AGN with $M_{BH}>10^{9}M_{\odot}$ where 
filled-in and open squares denote disk- and bulge-dominated hosts 
respectively. Horizontal lines denote the median IR luminosity of normal 
disk-dominated (upper) and bulge-dominated (lower) galaxies respectively (see 
\S\ref{sec:hockey} for discussion). Errors as in Fig.~\ref{fig:hockey}.
\label{fig:gr2}}
\end{figure}

In Fig.~\ref{fig:gr2} we plot mean observed 12$\micron$ IR luminosity versus 
mean observed 2-10keV X-ray luminosity for all Group 2 AGN at $z<0.1$, divided
 according to both estimated black hole mass and host galaxy classification. 
From Fig.~\ref{fig:gr2}, we note that Group 2 AGN in bulge-dominated hosts 
tend to be either less IR luminous \emph{or} have a lower (more Group 1-like) 
value of $R_{IR/X}$. 

\section{The hockey-stick shape distribution of AGN}
\label{sec:hockey}
We saw in Fig.~\ref{fig:hockey} that the distribution of the AGN in 
$R_{IR/X}=L_{IR}/L{X}$ is 'hockey-stick' shaped. The homogeneous Group 1
 AGN lie in a fairly tight range of luminosity ratio, forming the 'stick'. 
The heterogeneous Group 2 AGN flare out to much larger values of $R_{IR/X}$, 
forming the 'blade' of the hockey stick. In this 
section, we discuss the Group 2 'blade' of the hockey-stick distribution. 

There are two main causes of a Group 2 AGN blade in the 
hockey-stick distribution. First, unresolved IR emission from the host galaxy
(e.g. nuclear starburst or star formation) may dominate the AGN IR 
luminosity, pushing individual low IR luminosity AGN vertically up
 off the main (Group 1) distribution (the 'stick'). Contamination by non-AGN 
emission will therefore add a natural floor to the observed distribution of 
$L_{IR}$ in AGN. Second, the observed X-ray luminosity may be a small 
fraction of the intrinsic X-ray 
luminosity due to absorption, pushing individual highly absorbed AGN 
horizontally (leftwards to lower X-ray luminosities) off the main Group 1 
distribution (the 'stick'). A clear test of this possibility is to look at a 
sample of 
Group 1 and Group 2 AGN unbiased by photoelectric absorption, where we expect
 the absorbed Group 2 AGN 'blade' to collapse back onto the 'stick'. We carry
 out such a test and discuss the results in \S~\ref{sec:hardx} below. A third 
possibility, that changes in the AGN obscurer with $M_{BH}$ conspire to keep 
$L_{IR}$ flat with $M_{BH}$, can be ruled out. This is because an increase in 
bolometric luminosity with $M_{BH}$ (expected from the standard model and 
Fig.~\ref{fig:mass1}) should emerge as heating in the IR band. However we 
find that 
$L_{IR}$ remains flat with $M_{BH}$ across the IRAS 12-100$\micron$ bands. We 
shall now consider the contribution of unresolved emission to the formation 
of the Group 2 AGN 'blade'.

We saw in \S~\ref{sec:gr1} that Group 1 AGN in disk-dominated hosts have a 
larger mean observed IR luminosity than those in bulge-dominated hosts, 
probably due
 to contamination from the unresolved disk (see Fig.~\ref{fig:class1}). We 
should therefore expect to see something similar for the Group 2 AGN. In this 
vein, it is noteworthy that in Fig.~\ref{fig:class2}, the top of the Group 2 
'blade' tends to be occupied by AGN in disrupted hosts (green open circles) 
and the base of the 'blade' is mostly AGN in bulge-dominated hosts (red 
filled-in circles), with AGN in disk-dominated hosts (black crosses) dispersed
 between the two. The quantity of dust and the star formation rate in galaxies
 declines in the order: disrupted, disk-dominated, bulge-dominated. Therefore 
the average unresolved host contribution should indeed decrease going from 
disrupted to disk-dominated to bulge-dominated hosts. Since the 30'' 
IRAS beamwidth does not resolve the host galaxy in most cases, the observed IR
 luminosity could be dominated by the unresolved host contribution. 

We tested this possibility by comparing the the Group 2 'blade' of the hockey 
stick with the IR luminosity expected for normal bulge- and disk-dominated 
galaxies. From a very small recent sample of normal 
bulge-dominated galaxies \citep{b51} we found a median 12$\micron$ 
IRAS luminosity of $0.26 \times 10^{42}$ erg $\rm{s}^{-1}$, which is slightly
 below the median 
$L_{IR}$ of the bulge-dominated Group 2 AGN (red filled-in circles) in 
Fig.~\ref{fig:class2}. However, a much larger (and statistically 
more reliable) sample of 148 normal elliptical galaxies \citep{b94} yields a 
higher mean (median) IRAS 12$\micron$ luminosity of $2.40(2.63) \times 
10^{42}$ erg 
$\rm{s}^{-1}$, although some low luminosity AGN may be included in this 
estimate. This latter median luminosity is indicated in Fig.~\ref{fig:gr2} by 
the lower of the two horizontal dashed lines. In the 
case of disk-dominated galaxies, \citep{b93} find in a sample of 218 normal 
spiral galaxies ($T>0$) a mean(median) IRAS 12$\micron$ luminosity of 
$2.93(3.89) \times 10^{42}$ erg $\rm{s}^{-1}$. The $\sim 10^{42}$ erg 
$\rm{s}^{-1}$ difference between the median 12$\micron$ luminosity of 
elliptical and spiral galaxies is 
consistent with our rough estimate of a median IR contribution of 
$\sim 10^{42}$ erg $\rm{s}^{-1}$ from unresolved disks for Group 1 AGN (see 
\S~\ref{sec:gr1} above). Evidently these median luminosities exceed the IRAS
12$\micron$ luminosity of $\sim 3/4$ of the Group 2 AGN in 
Fig.~\ref{fig:hockey}. So, unresolved host galaxy emission could indeed 
dominate the IR luminosity of a majority of the Group 2 AGN in our sample. 
Furthermore, even recent looks with Spitzer at distant AGN are consistent 
with IRAS observations, indicating that it is difficult
 to separate the non-AGN and AGN IR components except for nearby objects 
\citep{b88}.

Fig.~\ref{fig:class2} shows that the Group 2 'blade' can be divided roughly  
between bulge-dominated hosts with generally lower $L_{IR}$ and 
disk-dominated hosts with higher $L_{IR}$. A generally lower $L_{IR}$ in 
bulge-dominated hosts is presumably due to the presence of less warm gas, 
dust and star formation than in disk-dominated hosts. This may be due in 
part to past episodes of intense AGN activity inhibiting star formation and 
blowing out gas and dust. We expect past episodes of feedback to have been 
greatest for AGN containing the largest mass black holes in the largest 
bulges. Sure enough \cite{b1} have found evidence that AGN at $0.05<z<0.1$ in 
elliptical host galaxies suppressed star formation in their hosts, and we 
discuss the relevance of this to our sample in \S\ref{sec:group2} below.

\section{Group 2 AGN}
\label{sec:group2}
We now have a relatively straightforward picture of the homogeneous Group 1 
AGN, where the 
central engine generally scales with black hole mass (Fig.~\ref{fig:mass1}). 
Our picture of the heterogeneous Group 2 AGN is far less straightforward. 
Group 2 AGN span a wide range of activity classification (Seyfert 2, LINERs, 
$\rm{H}_{II}$ regions as well as cross-classifications) and there could be 
many causes of their observed IR 
and X-ray luminosities. In this section, we study the Seyfert 2 AGN and 
LINER subsets of Group 2 AGN and we shall discuss models of their activity.

\subsection{Seyfert 2 AGN}
\label{sec:sy2s}
\begin{figure}
\includegraphics[height=3.35in,width=3.35in,angle=-90]{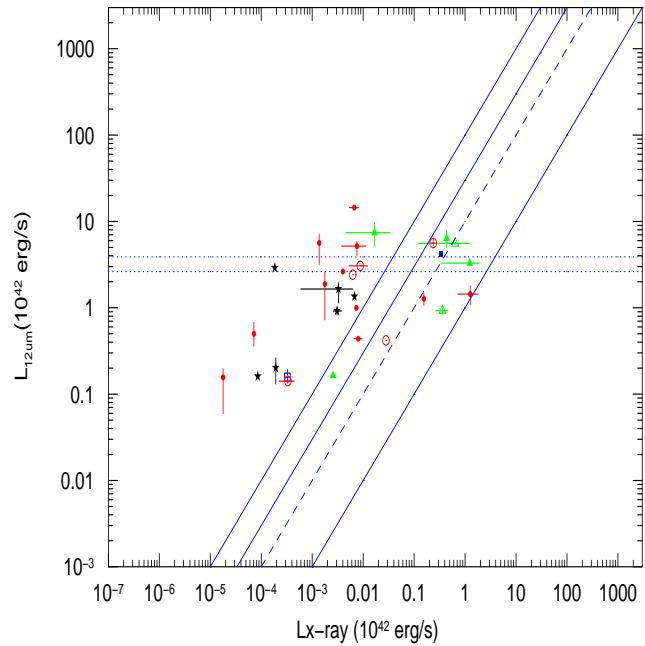}
\caption{As Fig.~\ref{fig:gr2}, except we plot Seyfert 2 AGN only. 
\label{fig:sy2s}}
\end{figure}

In the standard model, Seyfert 2 AGN are Seyfert 1 AGN seen edge-on to 
the observer's sightline \citep{b2,b90}. In Fig.~\ref{fig:sy2s} we plot only 
those objects from Fig.~\ref{fig:gr2} classified or cross-classified as 
Seyfert 2 AGN. From Fig.~\ref{fig:sy2s}, 
$L_{X}$ tends to increase with $M_{BH}$, particularly if we separate out 
low $M_{BH}$ AGN (black points, $<10^{7}M_{\odot}$) and high $M_{BH}$ AGN 
(green and blue points, $>10^{8}M_{\odot}$). However, $L_{IR}$ remains flat 
with large scatter, therefore $R_{IR/X}$ tends to decrease with $M_{BH}$. If 
Seyfert 2 AGN are edge-on versions of Seyfert 1 AGN 
(Group 1), then we should expect \emph{intrinsic} $L_{X}$ and $L_{IR}$ to 
broadly increase with $M_{BH}$ as in Fig.~\ref{fig:mass1}. The fact that 
\emph{observed} $L_{X}$ does seem to increase with
 $M_{BH}$ indicates that X-rays do escape from the AGN, presumably via 
scattering, reflection or leakage through variable covering, but this is a 
small fraction of the expected intrinsic $L_{x}$ (see e.g. the Group 1 AGN in 
Fig.~\ref{fig:mass1}).

If Seyfert 2 AGN are Seyfert 1 AGN seen edge-on, we also
 expect observed $L_{IR}$ to increase on average with $M_{BH}$, which we do 
not observe in Fig.~\ref{fig:sy2s}. $L_{IR}$ remains flat (with large 
scatter) for these Seyfert 2 AGN at longer IRAS wavelengths so the increased 
bolometric luminosity is not being reprocessed and emitted in the IRAS 
bandpass. Instead it seems likely that that the IRAS beamwidth ($30''$) 
introduces a 
floor to IR luminosity, due to emission from the unresolved host galaxy as we 
discussed in \S~\ref{sec:hockey} above. Indeed unresolved emission remains a 
problem even for modern high resolution IR spectroscopy of nearby AGN 
\citep{b16}.

\citet{b52} find higher ($0.9$ mag $\rm{arcsec}^{-2}$) median K-band 
(2.2$\micron$) surface brightness of disks in Seyfert galaxies compared to 
normal spiral galaxies. If the disks of Seyfert AGN 
hosts are significantly brighter than normal spiral disks at 12$\micron$, 
then based on the unified model of AGN, Seyfert 2 AGN in disk-dominated hosts 
should lie near the top of the Group 2 'blade' (near 
$\sim 10^{43}$ erg $\rm{s}^{-1}$) of the hockey-stick distribution. However, 
from Fig.~\ref{fig:sy2s}, only Seyfert 2 AGN with $M_{BH}>10^{8}M_{\odot}$ 
(green and blue points) have 
higher median $L_{IR}$ than for normal spiral galaxies (upper horizontal 
line).

Note that the median $L_{IR}$ of the 22/30 disk-dominated Seyfert 2 AGN in 
Fig.~\ref{fig:sy2s} is $\sim 1.4 \times 10^{42}$ 
erg $\rm{s}^{-1}$ which, for a small sample, remains consistent with the 
median $L_{IR}$ of a normal spiral galaxy ($\sim 3.9 \times 10^{42}$ erg 
$\rm{s}^{-1}$ see \S\ref{sec:hockey} 
above). The 8/30 Seyfert 2 AGN in bulge-dominated hosts in 
Fig.~\ref{fig:sy2s} tend to have lower a luminosity ratio range ($R_{IR/X}$) 
than their disk-dominated counterparts. Future high angular resolution IR 
studies of AGN could reveal Seyfert 2 AGN uncontaminated by host galaxy 
emission \citep{b97}. However, we can also test the hypothesis that Seyfert 2 
AGN have the same central engine as Seyfert 1 AGN by contrasting the very 
hard X-ray (14-195keV) emission for Group 1 and Group 2 AGN (see 
\S\ref{sec:hardx} below).

\subsection{LINERs}
\label{sec:liners}

LINERs are a low-ionization sub-set of galactic nucleus activity. It is 
unclear whether LINERs are due to a different type of accretion flow onto the 
SBH, an evolutionary state of AGN activity, or merely a low accretion rate 
version of regular AGN activity. \citet{b89} find that LINERs 
(at $z>0.04$) occur in more massive, older hosts 
with higher stellar velocity dispersions and that there is a 
smooth transition from Seyfert AGN to LINER, suggesting that LINERs are 
indeed (mostly) AGN. \citet{b1} suggest that LINERs are likely 
transitioning from more active to less active states. \cite{b1} also found 
evidence that AGN at $0.05<z<0.1$ in elliptical host galaxies suppressed 
star formation which is relevant to our discussion below. 

\begin{figure}
\includegraphics[height=3.35in,width=3.35in,angle=-90]{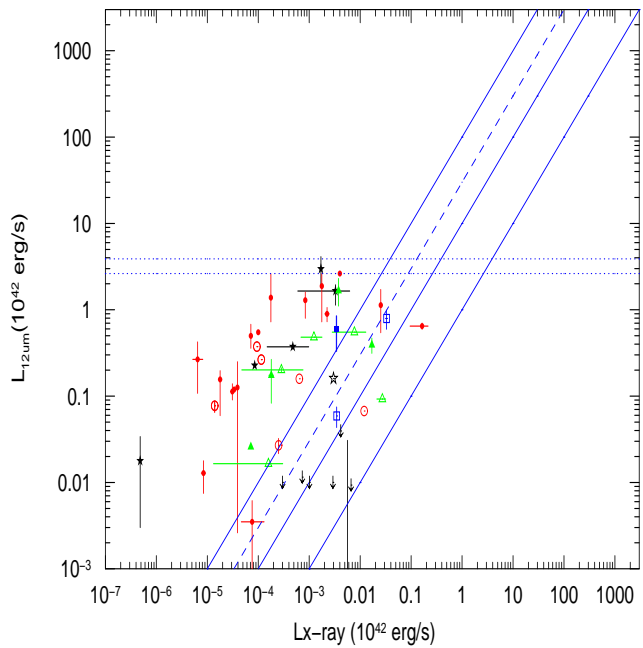}
\caption{As Fig.~\ref{fig:gr2}, except for AGN with LINER classifications. 
Symbols and colours are as in Fig.~\ref{fig:gr2}. 
Also included are measurements for ULXs from \citet{b18}, denoted by downward 
pointing arrows (upper limits based on IRAS non-detection), see discussion in 
\S\ref{sec:ulx}.
\label{fig:liners}}
\end{figure}

In Fig.~\ref{fig:liners} we plot only those objects from Fig.~\ref{fig:gr2} 
that are classified or cross-classified as LINERs. From 
Fig.\ref{fig:liners}, the median observed $L_{IR}$ for LINERs in 
disk-dominated hosts is well below the median for normal spiral galaxies 
(upper horizontal dashed line) suggesting either older stellar populations 
in the hosts or past prodigious feedback in both bulge- and 
\emph{disk-dominated} hosts of present-day LINERs. Interestingly, the 
disk-dominated LINER hosts in our sample tend to lie close by 
($<30$Mpc or $z<0.01$). This may indicate that disk-dominated LINERs have 
been undercounted in the
 SDSS ($z>0.04$) surveys \citep{b89}, possibly due to lower than average 
surface brightness of the hosts or it may be that some of the disk-dominated 
LINERs are misclassified. Recently \citet{b21} suggested that some LINERs 
might correspond to 'retired' galaxies,
 where ionizing radiation originates in post-AGB stars and white dwarfs 
rather than a weakly accreting black hole. 
This is an intruiging possibility, but it begs the question, if 
the X-ray luminosity in these sources does not come from accretion onto a 
supermassive black hole, where does it come from?

\subsection{Could ULXs live in galactic nuclei?}
\label{sec:ulx}
Not all nuclear activity may be due to accretion onto a
 supermassive black hole. \citet{b5} propose that LINER 2s (LINERs without 
broad $\rm{H} \alpha$ wings) with no flat-spectrum compact radio 
cores are not AGN at all, but rather 'heterogeneous' in origin. If 
radio-quiet LINER 2s are not due to accretion onto a supermassive black 
hole, then what produces the narrow optical lines and X-ray emission? Recent 
studies suggest that post-AGB stars and white dwarfs could account for the 
ionization in some LINERs \citep[e.g.][]{b21,b22}. Narrow 
optical lines in such low luminosity nuclei can certainly also be produced as 
a result of star formation \citep{b5}, but what about the X-ray emission and 
estimates of a large central black hole mass?

A large mass estimate for the central $M_{BH}$ from e.g. bulge properties, 
combined with nuclear hard X-ray continuum does not automatically imply AGN 
activity for these radio quiet LINER 2s. A quiescent supermassive black hole 
could exist at the center of the host galaxy, leading to an (appropriate, 
accurate) large $M_{BH}$ 
estimate. Furthermore, there are several candidates capable of producing the 
observed X-ray luminosities. X-ray luminosities $<10^{39}$ ergs $\rm{s}^{-1}$ 
can be produced by high mass X-ray binaries, known as galactic black hole 
candidates (GBHCs) in our Galaxy. GBHCs are observed to reproduce all the 
aspects of an X-ray spectrum of an AGN, including broad and narrow Fe K 
emission 
\citep{b7,b11} and reflection features. GBHCs also transition from soft, 
high states to hard, low states on short timescales, which may be analagous to
 the Seyfert and LINER states in AGN on much longer timescales 
\citep[e.g.][]{b89}. X-ray luminosities $10^{39}<L_{x}<10^{42}$ ergs 
$\rm{s}^{-1}$ can be produced by ultra-luminous X-ray sources (ULXs) 
\citep[e.g.][]{b34}. ULXs are powered by either large stellar mass black holes 
($>20 M_{\odot}$) accreting at near Eddington luminosities (possibly with 
beaming) or intermediate mass black holes (IMBHs) accreting at a fraction of 
their Eddington luminosities. ULXs are defined as non-nuclear X-ray sources, 
but there is no reason to expect that a large stellar mass black hole or IMBH 
should not live in the nucleus of a galaxy, particularly if there have been 
periods of enhanced star formation. So, we 
propose that if these 'radio-quiet' LINER 2s are not actually AGN, then 
their X-ray emission originates in nuclear ULXs. In the case of LINER 2
 nuclei without flat-spectrum, compact radio cores, a large $M_{BH}$ 
estimate and detection of a hard X-ray continuum/ Fe K emission may not 
actually indicate AGN activity. There is a simple test of this model for 
nuclear ULX emission, namely a combination of short timescales of X-ray 
variability and a large fractional
X-ray variability (excess variance) combined with spectral changes. X-ray 
variability timescales alone are 
insufficient, since partial covering changes could account for very small 
timescales \citep[e.g.][]{b4,b3}. However, very large fractional variability 
combined with a change in the spectrum indicate a large stellar mass black hole
with GBHC-like behaviour. If an IMBH is responsible for the X-ray emission, 
then the fractional X-ray variability should be intermediate between AGN and 
GBHCs and the black-body 
temperature of the accretion disk should be lower (in the UV to soft X-ray 
band) than for a GBHC (soft X-ray band) \citep{b34}. Furthermore, if ULXs are 
responsible for the X-ray emission in these galactic nuclei, then the low 
ionization optical radiation must originate somewhere other than accretion 
onto a supermassive black hole. One intruiging possibility is that these 
galaxies may have retired from star formation and therefore produce hotter 
emission line regions \citep{b21}. Optical studies of 'radio-quiet' LINER 2 
host galaxies using SDSS could test the reality of LINER activity in these 
hosts.

Fig.~\ref{fig:liners} includes several ULXs (denoted by 
downward pointing arrows) from \citet{b18}. Note 
that the 2-10keV luminosity estimates for the ULXs are based on 
model fits rather than observed flux. The ULXs have upper limits to 
$L_{IR}$ (based on the IRAS flux detection limit and estimated distance). 
Based on the upper limits of $R_{IR/X} \sim 30$, the ULX 
luminosity ratio is more akin to Group 1 AGN than Group 2 AGN. If $L_{IR}$ for
 the ULXs is considerably lower than the upper limits depicted in 
Figs~\ref{fig:gr2} and ~\ref{fig:liners}, then $R_{IR/X}$ for 
ULXs could be very small ($<1$), implying that the ULXs are jet-dominated or 
strongly beamed \citep[see][]{b99}. New observations of ULXs by Spitzer (yet 
to be published at time of writing) will hopefully reveal interesting limits 
on their IR luminosity.

\section{Testing models of AGN activity driven by host galaxy interactions}
\label{sec:rings}
External tidal interactions have 
long been thought of as one possible fuelling source for AGN \citep{b23} and 
certainly mergers and interactions may have been powerful drivers of nuclear 
activity in earlier epochs. Interactions may produce some nuclear activity in 
the form of a starburst, which 
may in turn lead to later AGN activity \citep{b25}. However, there are many problems 
with this model of fuelling AGN in the local Universe. It has been
 known for a long time that optically identified AGN occur far more frequently
 in field galaxies ($\sim 5\%$) than in clusters ($\sim 1\%$) \citep{b70}, 
which suggests that either (1) higher rates of interaction somehow 
inhibit AGN fuelling, (2) interactions are not significant drivers of AGN 
activity in general or (3) AGN activity lags the actual interaction by a very
 long time. Recent studies suggest that mergers are not a significant driver 
of optically selected AGN activity \citep[e.g.][]{b76,b50}, see however 
\citet{b6}. The internal 
properties of galaxies and feedback seem to have a greater correlation with 
optically selected AGN activity than density of galaxies in the host galaxy 
environment \citep{b26,b1}. Using X-ray selected 
AGN, the story is not very different. The AGN fraction drops going from 
sparse groups to dense clusters, even as the rate of interaction or merger 
should be increasing \citep{b75}, confirming optical results. Among AGN in 
compact groups the broad-line region may be entirely absent, possibly due to 
a \emph{decrease} in gas flow to the nucleus as a result of 
mergers/interactions \citep{b92}. Among X-ray
 selected AGN in disk dominated galaxies, internal instabilities or at most, 
minor interactions appear to be the key driver of AGN fuelling \citep{b77}. 
Indeed studies of 
the number of companion galaxies around X-ray selected AGN suggest that strong
 interactions occur at the same rate for AGN hosts as for normal galaxies 
\citep{b78}. Among X-ray and IR selected AGN, host galaxies are more likely to
 be closely paired than isolated, however the pairs are found to be 
non-interacting and undisturbed \citep{b79}. So, if interactions are actually 
important in fuelling AGN, observations suggest a non-trivial dependency and 
it is unclear exactly how the link can be made. 

AGN fuelling requires dust and gas in the central $\sim$ pc of a galaxy. Since
 most gas and dust in a galaxy lives far from the galactic nucleus, it 
is unclear how the gas can lose angular momentum and migrate inwards to fuel 
AGN. Several mechanisms have been proposed, but at best there must be a 
hand-off between different mechanisms operating on different distance- and 
size-scales \citep{b74}. In this section, we point out that AGN host galaxies 
with deVaucouleurs outer ring (R,R$^{\prime}$) morphological classification 
have been overlooked as a very useful probe of models of host interaction 
leading to AGN activity. Since Lindblad resonance rings seem to be very 
sensitive to small changes in the host galaxy \citep[e.g.][]{b31,b73} it turns
 out that they are a particularly powerful test of models of AGN fuelling via 
interactions and mergers.

Ringed structures are relatively common in and around galaxies in the nearby 
Universe, with three main varieties: collisional and polar rings (which we 
shall not discuss further here) and diffuse Lindblad resonance rings 
\citep[see e.g.][for a review]{b31}. Resonance rings are believed to form via 
the actions of galactic bars and can take a long time to form ($>3$Gyrs for 
some outer rings) \citep{b31}, $\sim$Gyr for certain pseudo-rings (R$^{\prime}$) 
\citep{b73}. It is hard to distinguish between ring varieties for distant 
galaxies, but candidate resonance rings at $z\sim 0.8$ could even take 
$\sim 6$Gyrs to form \citep{b80}. Resonance rings in galaxies have long been 
considered fragile, easily destroyed by tidal interactions \citep{b31}, or 
even by minor changes in bar pattern speed \citep{b73}. This fragility, 
apparent from simulations, makes resonance ring structures around galaxies a 
particularly useful probe of the relatively recent interaction history of a 
galaxy.

In our sample, we find that 9/24 Group 1 AGN and 20/97 Group 2 
AGN in ($T\geq -3$) host galaxies $<70$ Mpc distant, possess outer 
(R,R$^{\prime}$) rings \citep{b71}. This rate is consistent with the 
occurrence of (R,R$^{\prime}$) in the IRAS 
sample of galaxies \citep{b10}. These AGN host galaxies and their (ringed) 
de Vacouleurs classification are listed in Table~\ref{tab:rings}. There are 
ringed 
host galaxies in our sample out to $\sim 220$Mpc, but ringed host galaxies are
 undercounted if the rings have low surface brightness or are edge-on, so our 
selection is naturally biased by distance. Thus we limit ourselves to 
analyzing the sample of ringed host galaxies at $<70$Mpc following 
\citet{b10}. The RS and T-statistic tests 
reveal that the AGN hosts with outer rings (R,R$^{\prime}$) in 
Table~\ref{tab:rings} are drawn from the same population of 
$L_{IR}$,$L_{X}$ and $R_{IR/X}$ as the rest of the Group 1 and Group 2 AGN 
(within 70Mpc) at probabilities of $\sim 15-90\%$. Therefore, since galaxies 
with outer rings 
have presumably had no tidal interactions in \emph{at least} the past 
$\sim$Gyr (see above), present AGN activity (which lasts few $\times$10Myrs) 
is very unlikely to be due to interactions (NGC 2685 is the possible 
exception since it has an outer ring but is also cross-classified as 
peculiar). Furthermore, if tidal interactions generated the activity observed
 in our control group (no rings, $<70$Mpc), it is indistinguishable from 
activity due to secular processes. Note that we can conclude the same thing 
from the Group 1 AGN back in Fig.~\ref{fig:mass1}, where AGN in disk-dominated 
and (presumably tidally) disrupted hosts have a very similar spread in 
$L_{IR}$ and $L_{X}$. The RS and T-statistic 
tests on AGN at $<70$Mpc also reveal no 
significant difference in $M_{BH}$ between those host galaxies with and 
without outer rings (R,R$^{\prime}$), when we take into account uncertainty in
 the mass estimates. Therefore it seems likely that the AGN in our sample 
have had approximately their present $M_{BH}$ for a long time as suggested by 
recent simulations (e.g. \citet{b24,b43,b42,b41}). If tidal interactions 
are not the main driver of AGN activity in
 our sample, this suggests a number of possible explanations. The most likely
 explanation is that the timescale for activity driven by 
processes internal to the galaxy ($t_{int}$) is much smaller than the timescale
 for activity driven by processes external to the galaxy ($t_{ext}$), or
 $t_{int}\ll t_{ext}$. It is easy to see that estimates
 of rapid bar dissolution times and galaxy rotation speeds \citep{b73} are 
much shorter than the expected dynamical timescales of tidal interactions.

\begin{table}
\begin{minipage}{60mm}
\caption{AGN in our sample within 70Mpc possessing outer ring (R,R$^{\prime}$) 
deVacouleurs classifications \citep{b71}.
\label{tab:rings}}
\begin{tabular}{@{}lrrrr@{}}
\hline
Galaxy & Classification\\
\hline
Group 1 &\\
\hline
Mkn 766  & (R$^{\prime}$)SB(s)a\\
NGC 3516 & (R)SB(s)$0^{0}$\\
NGC 3783 & (R$^{\prime}$)SB(r)ab\\
NGC 4151 & (R$^{\prime}$)SAB(rs)ab\\
NGC 4593 & (R)SB(rs)b\\
NGC 5548 & (R$^{\prime}$)SA(s)0/a\\
NGC 6860 & (R$^{\prime}$)SB(r)ab\\ 
NGC 7469 & (R$^{\prime}$)SAB(rs)a\\
ESO323-G077 & (R)SAB(rs)$0^0$\\
\hline
Group 2 & \\
\hline
NGC 1068 & (R)SA(rs)b\\
NGC 1808 & (R)SAB(s)a\\
NGC 2639 & (R)SA(r)a\\
NGC 2681 & (R$^{\prime}$)SAB(rs)0/a\\
NGC 2685 & (R)SB0+;pec\\
NGC 3185 & (R)SB(r)a\\
NGC 4457 &(R)SAB(s)0/a\\
NGC 4507 & (R$^{\prime}_{1}$)SAB(rs)b\\
NGC 4736 & (R)SA(r)ab\\
NGC 4750 & (R)SA(rs)ab\\
NGC 4941 & (R)SAB(r)ab\\
NGC 4968 & (R$^{\prime}$)SAB$0^0$\\
NGC 5347 & (R$^{\prime}$)SB(rs)ab\\
NGC 7465 & (R$^{\prime}$)SB(s)$0^{0}$\\
NGC 7582 & (R$^{\prime}$)SB(s)ab\\
MCG-5-23-16 & (RL)SA(l)$0^{0}$\\
Mkn 573 & (R)SAB(rs)0+\\
Mkn 1066 & (R)SB(s)0+\\
Mkn 1157 & (R')SB0/a\\
IC 4709 & (R$_{1}$)SB(r)0+\\
\hline
\end{tabular}
\end{minipage}
\end{table}

For galaxies in the IRAS sample within 70Mpc, outer ring structures (not 
collisional) are found
 in only $\sim 10\%$ (48/479) of normal galaxies, but are much 
more common in Seyfert AGN host galaxies (19/57) \citep{b10}. An intruiging 
possibility is that tidal interactions 
actually disrupt the action of bars so that both ring formation and AGN 
activity are inhibited, which would be consistent with conclusions from 
recent optical and X-ray studies of AGN \citep{b92,b75}. A simple test of 
this hypothesis is that the 
occurrence of AGN in recently disrupted barred spiral hosts should be lower 
than the occurrence of AGN among all barred spiral hosts. We shall return to 
this hypothesis in future work.

\section{Biases in our Sample}
\label{sec:bias}  
In \citet{b99} we discussed biases in our sample and our results, particularly
 with reference to the all-sky IRAS (12-100$\micron$) and soft X-ray (ROSAT) 
surveys. We pointed out that in a heterogeneous sample such as ours, there are
 many biases, generally in favour of more luminous, local AGN, and against 
less luminous, more distant AGN. Here we briefly review some of the more 
important biases, but see e.g.\citet{b95} for warnings about 
heterogeneous samples of AGN in general and \citet{b99} for more detailed 
discussion of most of the present sample. 

On one hand, since we consider the 
reported observed AGN luminosity, we are biased against highly obscured AGN 
(particularly at larger distances). We also introduce error into luminosity 
estimates, since 
non-AGN host galaxy contributions to the luminosity are included in the 
observed luminosity in most cases. A further error in 
luminosity estimates is introduced since internal and Galactic absorption 
corrections are 
not systematically reported for all sources. On the other hand, an advantage 
of using observed luminosity is that we avoid assumptions about the central 
engine of AGN in order to estimate the intrinsic AGN luminosity. Furthermore, 
all but the highest resolution observations of the closest AGN include non-AGN 
contributions. By using the observed AGN luminosity, we avoid the difficult 
(and model-dependent) problem of de-coupling the non-AGN contribution (which 
may be significant in many lower-luminosity AGN in our sample). So, all our 
luminosity estimates include a non-AGN component of variable size. For 
example, in the X-ray band, the non-AGN component may consist of hot, diffuse
 gas, X-ray binaries or ULXs in the host galaxy. In the IR band, the non-AGN 
component may consist of star-forming regions or warm dust in the host 
galaxy. 

Our heterogeneous sample also introduces an 
obvious observation bias; many AGN have been observed in the 2-10keV X-ray 
band but were not seen with IRAS (or at least had upper limits). Likewise, 
some IRAS-detected AGN have not been observed in the 2-10keV band (or only 
have upper limits). None of these AGN have been included in our sample. In the 
case of the former (X-rays but no IR), we introduce a bias against AGN with 
very little warm dust 
(perhaps colder or more distant tori, or torus-less AGN with no star 
formation, or AGN with past episodes of feedback that destroyed small dust 
grains?). In the case of the latter (IR but no X-rays) we introduce a bias 
against sources with very low luminosity (radiatively inefficient?) accretion 
onto the central black hole in a relatively warm and dusty environment. 

In dividing our sample into two groups, we split according to optical line 
widths (the Seyfert 1.X) classification. However, for example, a Seyfert 1.5 
that is also classified as a starburst would be a Group 2 object in our 
sample. In this case, we introduce a bias in broad line 
objects against cross-classification (which may be due to obscuration, 
viewing angle or unrelated nuclear 
activity). ULIRGs are another good example of complexity in a source, since 
some are cross-classified as Seyfert, starburst and interacting objects and 
the fraction of their bolometric luminosity that is due to AGN can vary from 
$\sim 15-75\%$ \citep{b88}, making these obvious Group 2 candidates in our 
sample, with a potentially large non-AGN component. However, 
cross-classified ULIRGs are a very small fraction of our sample. For further 
detailed discussion of the X-ray and IR biases that are present in our sample,
 see \citep{b99}.
 
\subsection{Comparison with an X-ray survey unbiased by absorption}
\label{sec:hardx}
While there has been no all-sky survey in the 2-10keV band, new 
all-sky surveys are emerging in the hard ($>$10keV) X-ray band, unbiased by 
photoelectric absorption. Where X-rays are unbiased by absorption, we should 
be able to detect the \emph{intrinsic} X-ray luminosity of Group 2 AGN. This 
means that in a luminosity ratio plot analagous to Fig.~\ref{fig:hockey} using
 very hard X-rays, we 
should expect the 'blade' of the 'hockey stick' shaped distribution of AGN to 
collapse onto the 'stick'. In this section, we shall 
investigate the significance of our results by considering just those AGN in 
the hard X-ray all-sky surveys.

\begin{figure}
\includegraphics[height=3.35in,width=3.35in,angle=-90]{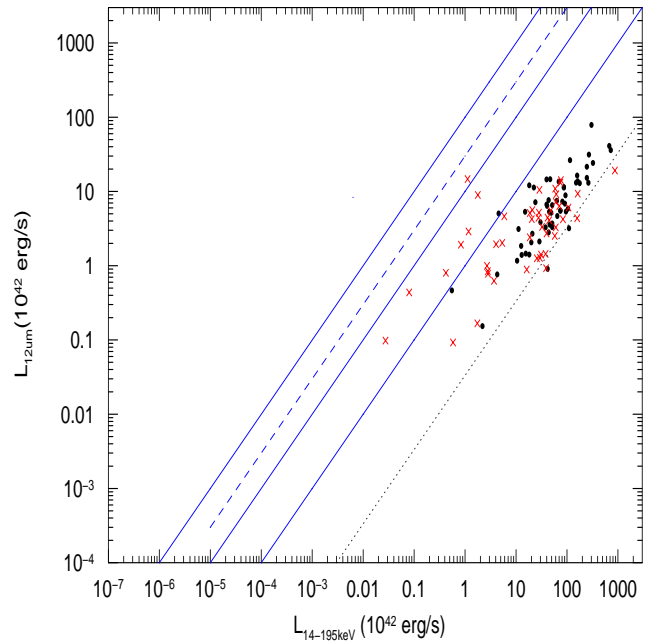}
\caption{12$\micron$ IRAS observed luminosity versus the very hard X-ray 
(14-195keV) luminosity for the 104 AGN in \citep{b57} with 12$\micron$ data. 
83/104 AGN are in our sample. Black filled-in circles are Group 1 AGN and red 
crosses are Group 2 AGN and the dotted black line indicates a luminosity ratio
 of $R_{IR/X}=1/30$.
\label{fig:batsample}}
\end{figure}

The 22-month, all-sky, \emph{SWIFT} BAT survey has a detection sensitivity of 
$2.2\times 10^{-11}$ erg $\rm{sm}^{-2} \rm{s}^{-1}$ over most of the sky in 
the 14-195keV band \citep{b57}. The 22-month BAT catalogue 
contains a total of 461 sources, including 262 AGN \citep{b57}, of which 228 
are 
Seyfert AGN and LINERs (equivalent to AGN in our sample). Of these AGN, 89 
were not detected with IRAS or had upper limits only at 12$\micron$. A further
 40 are either unidentified with a counterpart or have no 2-10keV flux 
measurements. The remaining 99 AGN are in our full sample. Of these 99 AGN, 
some 83 can be classified as Group 1 or Group 2 (the remaining 16/99 are 
highly beamed according to their NED classifications). In \citet{b57} 104 
AGN have both 12$\micron$ and 14-195keV luminosity. In 
Fig.~\ref{fig:batsample} we plot the very hard X-ray luminosity (14-195keV) 
from \citep{b57} versus mean observed 12$\micron$ IRAS luminosity for these 
104 AGN observed with BAT (83 of 
which are in our sample). Symbols are as in previous figures and the dotted 
line in Fig.~\ref{fig:batsample} represents $R_{IR/X}=1/30$.

Fig.~\ref{fig:batsample} reveals several things. First, the Group 1 AGN (black
 circles) have higher X-ray luminosities in the 14-195keV band than in the 
2-10keV band by a factor $\sim 30$ on average. This has the effect of moving 
the whole Group 1 distribution to the right in parameter space (since the 
12$\micron$ luminosity is obviously unchanged). Second, most of the Group 2 
AGN (red crosses) in Fig.~\ref{fig:batsample} lie on top of the distribution 
of Group 1 AGN. The 'hockey stick' distribution of Fig.~\ref{fig:hockey} has
collapsed onto the 'stick' in Fig.~\ref{fig:batsample}. This shows that the 
central engines of the Group 1 and most of
 the Group 2 AGN are identical (see also the discussion in \S\ref{sec:sy2s} 
above). Third, there is 
greater scatter among the Group 2 AGN. This is due to the more heterogeneous 
nature of the Group 2 AGN. In particular note that all 7/46 Group 2 AGN with 
$R_{IR/X}>1$ are classified wholly or in part as LINERs. These LINERs have 
hard X-ray luminosities one or two orders of magnitude less than the Seyfert 2 
AGN.  Fourth, the magnitude of the dispersion in the luminosity ratio of 55/56 
Group 1 AGN in Fig.~\ref{fig:batsample} is $\sim [1,1/30]$. This factor of 
$\sim 30$ in $R_{IR/X}$ dispersion in the 14-195keV band is identical to the 
dispersion range of $\sim 30$ for Group 1 AGN in the 2-10keV band 
(see Fig.~\ref{fig:hockey} above). In fact the dispersion in the 14-195keV 
band is slightly tighter since the factor of $\sim 30$ dispersion in 
$R_{IR/X}$ in the 2-10keV band applied to $\sim 90\%$ of Group 1 AGN. The fact
 that dispersion in $R_{IR/X}$ among Group 1 AGN is effectively constant 
between the 2-10keV and 14-195keV bands suggests that (a) the dispersion is 
intrinsic to the central AGN engine, apart from possible IR contributions from
 the host galaxy and (b) photoelectric absorption at 2-10keV for the Group 1 
AGN is insignificant.

\begin{figure}
\includegraphics[height=3.35in,width=3.35in,angle=-90]{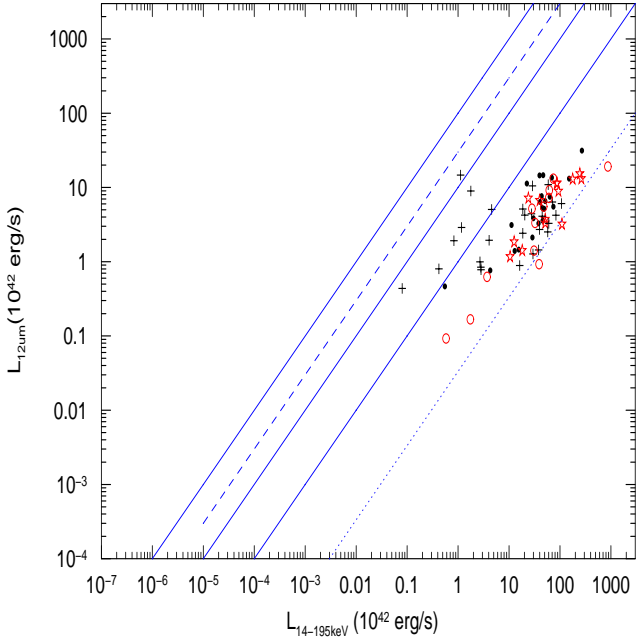}
\caption{As Fig.~\ref{fig:batsample}, except we divide the AGN according 
to host galaxy and AGN type. Red open stars are bulge-dominated Group 1 AGN, 
red open circles are bulge-dominated Group 2 AGN. Black filled-in circles are
 disk-dominated Group 1 AGN and black crosses are disk-dominated Group 2 AGN. 
AGN in peculiar or interacting hosts have not been plotted.
\label{fig:batsample_gal}}
\end{figure}

Figure~\ref{fig:batsample_gal} shows the AGN from Fig.~\ref{fig:batsample} 
divided according to host galaxy type. AGN in peculiar or interacting hosts 
have not been plotted. Fig.~\ref{fig:batsample_gal} shows a tight dispersion 
among AGN in bulge-dominated hosts (red open symbols). 95$\%$ of the AGN in 
bulge-dominated hosts (25/26) in Fig.~\ref{fig:batsample_gal} vary by a 
factor of $<9$ in $R_{IR/X}$. This compares well with the spread of $\sim 7$ 
in $R_{IR/X}$ for $\sim 90\%$ of the 
Group 1 AGN in bulge-dominated hosts in \S\ref{sec:gr1} above. 
The RS test shows that the bulge- and disk-dominated hosts (black symbols) 
in Fig.~\ref{fig:batsample_gal} have different mean values of $R_{IR/X}$ at 
$>99\%$ confidence. This difference is strongly driven by the (6) 
disk-dominated Group 2 AGN (black crosses) in Fig.~\ref{fig:batsample_gal} 
with $L_{X}<2 \times 10^{42}$ ergs $\rm{s}^{-1}$, all of which have nuclear 
starbursts which dominate or are comparable in luminosity to a weak AGN or 
LINER. If the intrinsic IR emission from these 6 AGN is much lower than 
observed they would drop vertically, possibly into the same 
distribution in $R_{IR/X}$ as the rest of the bulge-dominated AGN.

\begin{figure}
\includegraphics[height=3.35in,width=3.35in,angle=-90]{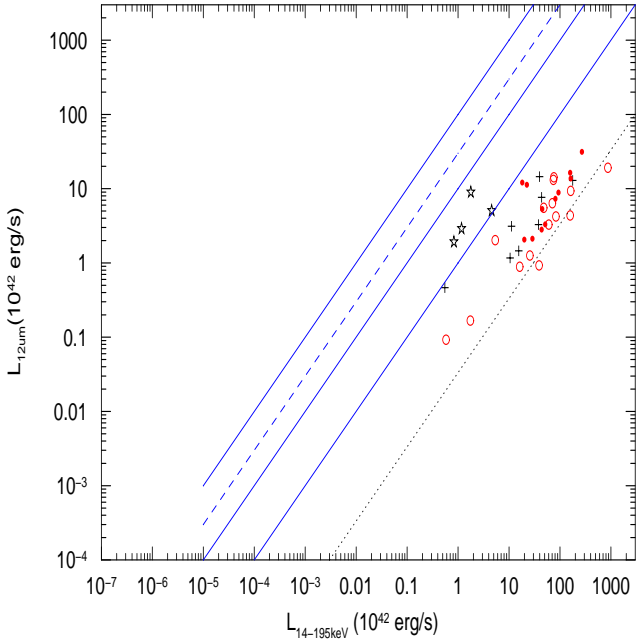}
\caption{As Fig.~\ref{fig:batsample}, except divided 
according to estimated central black hole mass. Black crosses denote low mass
($<10^{7}M_{\odot}$) Group 1 AGN. Black star symbols denote low mass 
Group 2 AGN.  Red filled-in circles denote high mass ($>10^{8}M_{\odot}$) 
Group 1 AGN and red open circles denote high mass Group 2 AGN.
\label{fig:batsample_mbh}}
\end{figure}

Figure~\ref{fig:batsample_mbh} shows the AGN from Fig.~\ref{fig:batsample} 
divided according to black hole mass. In order to
 minimize overlap due to measurement or systematic error, we divide 
AGN with black hole mass estimates into those $<10^{7}M_{\odot}$ (coloured 
black) and those $>10^{8}M_{\odot}$ (coloured red) (see also the discussion in
 \S\ref{sec:gr1} above). The RS test shows that all of the low mass 
($<10^{7}M_{\odot}$) black holes have a different mean X-ray luminosity (and 
$R_{IR/X}$) from the high mass ($>10^{8}M_{\odot}$) at 99.8$\%$ (99.2$\%$) 
confidence (or $\sim 3\sigma$ significance). This difference again appears to 
be driven by low mass, high $R_{IR/X}$, Group 2 AGN (black star symbols). We 
also investigated whether Group 1 and Group 2 
AGN in the 22-month BAT sample, within $70$Mpc, had host galaxies that were 
ringed (R,$\rm{R}^{\prime}$) or not. We found that there is no statistically 
significant difference in the X-ray luminosity or $R_{IR/X}$ between ringed 
and non-ringed AGN, although there was a difference in the mean IR luminosity
 of the ringed and non-ringed AGN hosts at the 95$\%$ confidence level which 
may have been driven by outliers.

In comparing our earlier results with those from an all-sky, very hard X-ray 
survey, we are testing the robustness of our conclusions with respect to X-ray
 sky coverage and photoelectric absorption. Because of the smaller size of the
 BAT sample, it is not possible to restate our conclusions with the same 
confidence, but it is noteworthy and comforting that the BAT sample supports 
our conclusions (albeit at lower confidence levels). In particular: (1) 
there remains a tight (model-independent) dispersion in $R_{IR/X}$ for 
(unobscured) AGN central engines, (2) AGN in bulge-dominated host galaxies 
have a narrower dispersion in $R_{IR/X}$ than in disk-dominated hosts, (3) 
across all Group 1 and Group 2 AGN there is a tendency towards higher mean 
X-ray luminosity and therefore lower $R_{IR/X}$ (for similar $L_{IR}$) with 
increasing black hole mass and (4) there is no 
statistically significant difference in $R_{IR/X}$ between ringed and 
non-ringed host galaxies. Our conclusions are 
robust at the level of sky coverage and completeness of the 22-month BAT 
sample.

\section{Conclusions}
\label{sec:conclusions}
We investigated the role of black hole mass ($M_{BH}$) and host galaxy 
morphology in a survey of the observed 2-10keV X-ray and 12$\micron$ IR 
luminosities of a heterogeneous sample of 276 mostly nearby AGN. We find 
that:\\

(a) As black hole mass increases, the average observed IR and X-ray 
luminosity of the homogeneous Group 1 (Seyfert 1.X) AGN increases,
 maintaining a ratio in the range $R_{IR/X}=$[1,30] over
 $\sim 3-4$ orders of magnitude in mass at a confidence level of $>90\%$. The 
luminosities for Group 1 AGN remain in the range $\sim 10^{-3}-10^{-1}$ of 
Eddington for each decade in mass. By 
contrast, among the heterogeneous, lower luminosity Group 2 AGN, the ratio 
$R_{IR/X}$ \emph{decreases} as black hole mass increases. This is mostly due 
to a combination of host galaxy contamination and photoelectric absorption. 

(b) There is an average increase in IR luminosity of a factor of $\sim 3$ 
among Group 1 AGN in disk-dominated hosts versus bulge-dominated hosts at a 
confidence level $>99\%$. Presumably the increase is due to contamination 
from the unresolved galactic disk \citep{b12}. Therefore a better measure of 
the underlying central engine is the range of $R_{IR/X}$ in bulge-dominated 
Group 1 AGN, which is $1<R_{IR/X}<7$ at $\sim$90$\%$ 
confidence. This dispersion 
incorporates \emph{all} possible variations among these Group 1 AGN, ranging 
from accretion mechanism and efficiency, to the opening angle of the disk and 
orientation to sightline, to covering fraction of absorbing material, to 
patchiness of the X-ray corona and even including variability. Group 1 AGN 
are quite remarkably identical. This strongly constrains all 
models of X-ray and IR production in AGN.

(c) Among the lower luminosity 
Group 2 AGN, those in bulge-dominated hosts tend to have substantially lower 
$L_{IR}$ and/or $R_{IR/X}$ compared to those in 
disk-dominated galaxies. Among LINERs we find much lower $L_{IR}$ than in 
normal disk-dominated or bulge-dominated galaxies representing possible 
evidence for past epochs of AGN feedback in both bulge and disk-dominated 
hosts.

(d) Within 70Mpc, AGN in host galaxies with outer Linblad resonance rings 
(R,R') are statistically indistinguishable in $L_{X},L_{IR}$ and $M_{BH}$ 
from AGN in host galaxies without rings. Since these rings are hard to form 
and easily destroyed by perturbations \citep{b31,b80,b73}, the ringed host 
galaxies are unlikely to have tidally interacted with companions for a long 
time.  Since AGN activity typically lasts a few $\times 10$Myrs, the most 
likely explanation for this is that the timescale for activity driven by 
processes internal to the galaxy ($t_{int}$) is much smaller than the timescale
 for activity driven by processes external to the galaxy ($t_{ext}$), 
or $t_{int}\ll t_{ext}$.

(e) \emph{ULXs in galactic nuclei could account for the observed X-ray 
luminosity in certain LINERs}. We propose that the X-ray 
emission from LINER 2 nuclei without flat-spectrum, compact radio cores 
\citep{b5} originates in ULXs (most likely large stellar mass black holes 
accreting at near Eddington luminosity). This hypothesis is testable by 
searching for a combination of short timescale X-ray variability, large 
fractional variability (the excess variance) and spectral changes in LINER 2 
nuclei without these radio properties. A large estimate for the central mass 
in this case may simply indicate the presence of a quiescent supermassive 
black hole. Our proposal may be consistent with the 'retired galaxies' of 
\citet{b21} and could be tested in that context in a study of the optical 
spectra of host galaxies of the 'radio quiet' LINER 2s.

(g) We tested results (a-d) above by comparing the 14-195keV luminosity and 
the 12$\micron$ IRAS luminosity of AGN in the BAT 22-month sample 
\citep{b57}. We conclude that photoelectric absorption is not important in the
 2-10keV band for Group 1 AGN, but it is important for many (though not all) 
of the Group 2 AGN and accounts for some of the 'hockey stick' shape of the 
distribution in Fig.~\ref{fig:hockey}.

In summary, comparing the observed 2-10keV X-ray and IR luminosities of AGN 
with 
estimates of the central black hole mass and host galaxy classification can 
provide a powerful diagnostic tool to constrain models of AGN activity, 
$M_{BH}$ and AGN feedback over cosmic time.

\section*{Acknowledgements}
We made extensive use of the NASA/IPAC 
Extragalactic Database (NED), operated by the Jet Propulsion Laboratory, 
CalTech, under contract with NASA. BM \& KESF gratefully acknowledge the 
support of the Department of Astrophysics of the American Museum of Natural 
History, PSC grant PSC-CUNY-40-397 and CUNY grant CCRI-06-22. Thanks to the 
referee for a useful report that helped improve the paper.


\label{lastpage}

\end{document}